\documentclass[prd,aps,preprint,nofootinbib,]{revtex4}
\usepackage{amssymb}
\usepackage{epsfig}
\usepackage{amssymb}
\usepackage{amsmath,amssymb,graphicx}
\usepackage{color}

\newcommand{\be}{\begin{equation}}
\newcommand{\ee}{\end{equation}}
\newcommand{\bea}{\begin{eqnarray}}
\newcommand{\eea}{\end{eqnarray}}
\def\lla{\left\langle}
\def\rra{\right\rangle}

\def\ssc{\scriptscriptstyle}
\def\lsim{\mathrel{\raise.3ex\hbox{$<$\kern-.75em\lower1ex\hbox{$\sim$}}} }
\def\gsim{\mathrel{\raise.3ex\hbox{$>$\kern-.75em\lower1ex\hbox{$\sim$}}} }

\begin{document}

%\draft
\preprint{{\vbox{\hbox{NCU-HEP-k050}
\hbox{Aug 2012}
\hbox{rev. Jan  2013}
\hbox{ed. Mar. 2013}
}}}
\vspace*{.5in}
%\twocolumn[\hsize\textwidth\columnwidth\hsize\csname
%@twocolumnfalse\endcsname

%\vspace*{.2in}
\title{Majorana versus Dirac mass from holomorphic supersymmetric Nambu-Jona-Lasinio model}

\author{ Yan-Min Dai }
%\email{gfaisel@cc.ncu.edu.tw}

\affiliation{
Department of Physics and %\\
Center for Mathematics and Theoretical Physics, National Central
University, Chung-Li, Taiwan 32054 }

\author{ Gaber Faisel }
\email{gfaisel@cc.ncu.edu.tw}

\affiliation{
Department of Physics and %\\
Center for Mathematics and Theoretical Physics,
National Central University, Chung-Li, Tawian 32054 and\\
Egyptian Center for Theoretical Physics, Modern University for
Information and Technology, Cairo, Egypt 11211 }

\author{ Dong-Won Jung }
\email{ dwjung@kias.re.kr}

\affiliation{
Department of Physics, National Tsing Hua University and \\
Physics Division, National Center for Theoretical Sciences,
Hsinchu, Taiwan, 300 and\\
School of Physics, KIAS, Seoul , Korea 130-722 }

\author{ Otto C. W. Kong }
\email{otto@phy.ncu.edu.tw}

\affiliation{Department of Physics and %\\
Center for Mathematics and Theoretical Physics, National Central
University, Chung-Li, Tawian, 32054 }

\vspace*{.5in}
%\begin{center}
%{\Large Abstract}
%\end{center}
\begin{abstract}
%\vspace*{.5in}
We study the theoretical features in relation to dynamical mass
generation and symmetry breaking for the recently proposed
holomorphic supersymmetric Nambu--Jona-Lasinio model. The basic
model has two different chiral  superfields (multiplets )with a
strongly coupled dimension five four-superfield interaction. In
addition to the possibility of generation of Dirac mass between
the pair established earlier, we show here the new option of
generation of Majorana masses for each chiral superfield. We also
give a first look at what condition may prefer Dirac over Majorana
mass, illustrating that a split in the soft supersymmetry breaking
masses is crucial. In particular, in the limit where one of the
soft masses vanish, we show that generation of the Majorana mass
is no longer an option, while the Dirac mass generation survives
well. The latter is sensitive mostly to the average of the two
soft masses. The result has positive implication on the
application of the model framework towards dynamical electroweak
symmetry breaking with Higgs superfields as composites.
\end{abstract}

%\date{\today}
\maketitle

%\pacs{xxxx}
\newpage

\section{Introduction}
Dynamical mass generation and symmetry breaking are very
interesting theoretical topics with important phenomenological
applications. One of the simplest models of the kind is the
Nambu--Jona-Lasinio (NJL) model \cite{Nambu:1961tp}. It is also
the first explicit model of spontaneous symmetry breaking. The
analysis of the nonperturbative gap equation  established that
with a strong enough four-fermion interaction, a symmetry breaking
Dirac fermion mass would  result. When applied to the electroweak
symmetry breaking of the Standard Model, it fails to give the
relatively {small} experimental top quark mass \cite{tSM}.
Introducing heavier fourth family quarks to take the role of the
top quark is essentially ruled out by other experimental
constraints. The more interesting option of supplementing
supersymmetry \cite{CLB,92} requires a too low $\tan\!\beta$ value
to stay phenomenologically viable. The latter situation can be
resolved with an alternative supersymmetrization of the NJL model
recently proposed \cite{034}. The version has a dimension five
four-superfield interaction, which otherwise mimics well the most
basic features of the NJL model.

Supersymmetry is an important theme in modern physics. One
especially attractive feature, in our opinion, is that the scalar
fields are now part of the chiral superfields with the chiral
fermions. The chirality forbids any gauge invariant mass before
breaking any symmetry. Moreover, the full matter (super)field
spectrum is now strongly constrained by the gauge symmetry and
their anomaly cancellation conditions. Introduction of the
vectorlike pair of Higgs superfields with their unnatural gauge
invariant mass in the usual formulation of the supersymmetric
Standard Model looks particularly unattractive from the
theoretical perspective. A NJL mechanism, with the Higgs
superfield(s) generated as composite and the electroweak scale
generated by strong dynamics is hence very appealing. The
holomorphic supersymmetric Nambu--Jona-Lasinio model (HSNJL)
\footnote{In the first effort to produce a supersymmetric version
of the NJL model, a dimension six  four-superfield interaction was
used \cite{BL}. It was found that soft supersymmetry breaking was
needed for the kind of model to produce dynamical symmetry
breaking and generate Dirac mass \cite{BL,BE}. The model, however,
lost the basic NJL model feature of having a single auxiliary
(super)field as  both the (quark) composite and the symmetry
breaking Higgs (super)field. The HSNJL model retains the latter
feature, but it has a holomorphic (dimension five) four-superfield
interaction which does not contain a four-fermion interaction. At
least to the extent that a chiral superfield is the
supersymmetrization of the fermion field, and the model has a four
(quark) superfield interaction inducing a composite of two (quark)
superfields with vacuum condensate to break symmetry and generate
Dirac superfield(fermion) mass, we consider it a
supersymmetrization of the NJL model. Terminology aside, the
physics features and their possible realization in nature are what
is interesting. }
 construction \cite{034,042} gives exactly
such a scenario that looks compatible with all known experimental constraints.
In Ref.\cite{042}, superfield gap equation analyses of the Dirac mass generation
have been performed for the HSNJL model and the old supersymmetric model. Nontrivial
symmetry breaking masses were established.

Distinguished from the old models, the HSNJL model is very rich in
interesting theoretical features, some of which we report here.
Firstly, the HSNJL model is also capable of generating Majorana
masses of the chiral superfields. Note that the basic model has
two superfields which could otherwise be the Dirac pair. In fact,
in the generic case, the story of dynamical mass generation and
hence the resulting symmetry breaking pattern becomes more
complicated than the naive Dirac mass generation analysis would
otherwise conclude. We present in this article an illustration of
the Majorana mass generation-a feature that is unknown for models
in the literature. Short of doing a comprehensive and fully
generic gap equation analysis, we will compare the Majorana mass
result here versus the Dirac mass result in our previous paper
\cite{042} to give a first answer to the competition of Majorana
versus Dirac superfield masses. We will discuss how a splitting
between the two input soft supersymmetry breaking masses favors
the generation of Dirac superfield mass. In a very interesting
particular case, we will show that at the limit, one of the soft
masses vanish, the nontrivial Majorana mass would be killed as the
nontrivial Dirac mass solution survives. Dynamical mass generation
also implies dynamical symmetry breaking in general, as discussed
in Ref.\cite{042}. The result has an interesting implication to
the application to electroweak symmetry breaking, though the focus
of the present paper is on the theoretical features.

The details of the calculations involved {are} very similar to
what we have presented in Ref.\cite{042}.  In the latter paper, we
succeeded in getting the gap equations from first principle
supergraph analyses both for the new HSNJL model and the old
supersymmetric NJL model\cite{BL,BE} with the result of the latter
case in perfect agreement with the one from the effective field
theory analysis \cite{BE} and that of the simple NJL limit. For
the interest of the general readers, we will only sketch the
analyses here by highlighting only the essential features.
Theorists interested in the details should also read carefully
Ref.\cite{042}. Further details together with a fully generic
comprehensive analysis of the HSNJL model will be presented in a
forthcoming publication \cite{next}. In particular, note that the
analyses rely heavily on the formulation of the generating
functional, mass parameters, and self-energy amplitudes as
superspace quantity as introduced in Ref.\cite{042}. That is, only
the full superspace analog of usual Minkowski spacetime field
theory.

\section{Dynamical Generation of Majorana Mass}
The basic model has two different chiral superfields (multiplets)
$\Phi_+$ and $\Phi_-$, presumably carrying different quantum
numbers. For instance, they may be different gauge multiplets. The
dimension five four-superfield interaction is given by \be
\label{d5} -\frac{G}{2} \int\!\! d^4 \theta \, \Phi_+\Phi_+
\Phi_-\Phi_- \, (1+ B \theta^2)\, \delta^2\!(\bar{\theta}) \;. \ee
It is really a superpotential term, as indicated by the
$\delta^2\!(\bar{\theta})$, hence holomorphic. In our earlier
works \cite{034,042}, the possibility of superfield condensate
$\lla \Phi_+\Phi_- \rra$ giving rise to a Dirac mass term
${\mathcal M} \Phi_+\Phi_-$ has been investigated. Unlike the
dimension six four-supefield interaction, the superpotential term
offers also the option of, say, a condensate of $\lla \Phi_+\Phi_+
\rra$ or $\lla \Phi_-\Phi_- \rra$ giving Majorana mass terms
${\mathcal M}_- \Phi_-\Phi_-$ and ${\mathcal M}_+ \Phi_+\Phi_+$,
respectively. To investigate the option, we proceed similar to the
Dirac mass analysis \cite{042} by deriving the gap equations for
${\mathcal M}_+$ and ${\mathcal M}_-$.

Consider the Lagrangian density written as
\bea
{\cal L}  &=&
%& &
\int\!\!  d^4 \theta ~
 \bigg[  \Phi_+^\dagger \Phi_+ (1-\Delta_+)
  + \Phi_-^\dagger \Phi_- (1-\Delta_-)
\nonumber \\ && + \big({\mathcal M}_+ \, \Phi_+ \Phi_+
\,\delta^2\!(\bar{\theta}) + {\mathcal M}_- \, \Phi_- \Phi_-
\,\delta^2\!(\bar{\theta}) + H.c. \big) \bigg]
%- m \Phi^{\dagger}_+ \Phi^{\dagger}_-\delta(\theta) %,\label{lzero}
\;\;+ \;\;{\cal L}_I \;,
\eea
with
\be
{\cal L}_I =   \int\!\! d^4 \theta ~
\bigg[- {\mathcal M}_+ \, \Phi_+^{ai} \Phi_+^{bi} \! - {\mathcal M}_- \,
\Phi_-^{\bar{a}j} \Phi_-^{\bar{b}j}\!
- \frac{G}{2} \Phi_+^{ai} \Phi_+^{bi} \Phi_-^{\bar{a}j}\Phi_-^{\bar{b}j} \,
 (1+ B \theta^2)\,
 \bigg] \delta^2\!(\bar{\theta}) + H.c.,
\nonumber \\
\ee where we consider explicitly $SU(N_c)\otimes SO(N_r)$
multiplets $\Phi_+^{ai}$ and $\Phi_-^{\bar{a}j}$ in the
fundamental and its conjugate representations of $SU(N_c)$ as
indicated by the $a$ and $\bar{a}$ (as well as $b$ and $\bar{b}$)
indices, \footnote{Note that we use here the $\bar{a}$ upper index
in place of a lower $a$ index with the summation convention
contracting it with the upper $a$ index.} %%
respectively, and the real fundamental representation of $SO(N_r)$
as indicated by the $i$ and $j$ indices. Moreover, in the first
part of the Lagrangian, we have suppressed all indices for
simplicity. The same applies to the expressions below; the indices
are not shown unless necessary. Note that the dimension five
interaction is invariant under the symmetry, while the Majorana
mass terms break the $SU(N_c)$ symmetry. For a clear comparison,
the Dirac mass term ${\mathcal M}\, \Phi_+^{ai}\Phi_-^{\bar{a}j}$
preserves $SU(N_c)$ symmetry while generally breaking the
$SO(N_r)$ symmetry.

Here, $\Delta_\pm = \tilde{m}^2_\pm \theta^2 {\bar \theta}^2$ characterizes
 the input  soft supersymmetry breaking mass-squared
$\tilde{m}^2_\pm$ for the corresponding scalar field $A_\pm$ and
${\mathcal M}_\pm$ superfield Majorana mass parameter \be
{\mathcal M}_\pm = m_\pm - {\theta}^2 \eta_\pm \;, \ee with the
supersymmetric Majorana mass $m_\pm$ and its supersymmetry
breaking counterpart $\eta_\pm$. The mass parameters in the above
equation are what we aim at generating dynamically. The gap
equations are given by \be \label{gap} - {\mathcal M}_\pm = \left.
\Sigma^{({\mbox \tiny loop})}_{\pm\pm}(p,\theta^2)
 \right|_{\mbox{\tiny on shell,}} \;
\ee where $\Sigma^{({\mbox \tiny loop})}_{\pm\pm}$ denotes the
lowest order contributions to the proper self-energy from loop
diagrams involving the four-superfield interactions. Note that
$\Sigma^{({\mbox \tiny loop})}_{++}$ has contribution involving
the $\Phi_-$ superfield propagator $\lla T(\Phi_-(1) \Phi_-(2))
\rra$, and $\Sigma^{({\mbox \tiny loop})}_{--}$ the propagator
$\lla T(\Phi_+(1) \Phi_+(2)) \rra$. The propagator should include
the Majorana masses ${\mathcal M}_\pm$ dependence. The propagators
are given in the same form as the Dirac case of $\lla T(\Phi_+(1)
\Phi_-(2)) \rra$, namely, as \bea \lla T(\Phi_\pm(1) \Phi_\pm(2))
\rra &=& \frac{i \, \bar{m}_\pm}{p^2(p^2+|m_\pm|^2)}
\frac{D_{\!\ssc 1}^2}{4} \delta^4_{\ssc 12} \nonumber \\&&
\hspace*{-.5in}
-\frac{i}{[(p^2+|m_\pm|^2+\tilde{m}^2_\pm)^2-|\eta_\pm|^2]} \!
\left[ \frac{\bar{\eta}_\pm \, D_{\!\ssc 1}^2  \bar{\theta_{\!\ssc
1}}^2}{4} - \frac{\eta_\pm |m_\pm|^2 \, D_{\!\ssc 1}^2
{\theta_{\!\ssc 1}}^2}{4p^2} \!\right] \!\delta^4_{\ssc 12}
\nonumber  \\&& \hspace*{-.5in} + \frac{i\, \bar{m}_\pm \;
[\tilde{m}^2_\pm (p^2+|m_\pm|^2
+\tilde{m}^2_\pm)-|\eta_\pm|^2]}{(p^2+|m_\pm|^2)[(p^2+|m_\pm|^2+\tilde{m}^2_\pm)^2-|\eta_\pm|^2]}
\; \left[\frac{D_{\!\ssc 1}^2 \theta_{\!\ssc 1}^2
\bar{\theta_{\!\ssc 1}}^2}{4} +  \frac{\bar{\theta_{\!\ssc 1}}^2
\theta_{\!\ssc 1}^2 D_{\!\ssc 1}^2}{4} \!\right] \delta^4_{\ssc
12} \;. \label{sprop} \eea We have the gap equations \bea m_\pm
&=& \frac{\bar{\eta}_\mp G N_r}{2}\;
I_2(|m_\mp|^2,\tilde{m}^2_\mp,|\eta_\mp|, \Lambda^2)\;,
\nonumber \\
\eta_\pm &=&  \bar{m}_\mp G N_r \;
I_1(|m_\mp|^2,\tilde{m}^2_\mp,|\eta_\mp|, \Lambda^2) -
\frac{\bar{\eta}_\mp G B N_r}{2} \;
I_2(|m_\mp|^2,\tilde{m}^2_\mp,|\eta_\mp|, \Lambda^2) \;, \eea
where $I_1(|m|^2, \tilde{m}^2, |\eta|, \Lambda^2) $ and
$I_2(|m|^2, \tilde{m}^2, |\eta|, \Lambda^2)$ are the same loop
integrals as before \cite{042}, the details of which we will
discuss below.

The first thing to note from the gap equation results is that they
are almost of exactly the same form as the Dirac case \cite{042}.
Actually, if we take identical soft masses
$\tilde{m}^2_\pm=\tilde{m}^2$, we have obviously a symmetric
solution relative to $\Phi_+$ and $\Phi_-$, which is exactly the
same as the gap equation for the Dirac mass case shown in
Ref.\cite{042} (with $N_r$ replaced by $N_c$).  For instance,
considering only the case of real values for $m_\pm$ and
$\eta_\pm$ under the assumption of a real and small $B$ value, we
find that a nontrivial solution exists for large enough $G$ (taken
as real and positive here by convention) satisfying \be G >
\sqrt{G_0^2 + b^2} + b \sim G_0  + b\;, \ee where \bea G_0^2 =
\frac{512\pi^2}{\tilde{m}^2\ln{\left(1+\frac{\Lambda^2}{\tilde{m}^2}\right)}
              \left[\ln{\left(1+\frac{\Lambda^2}{\tilde{m}^2}\right)}
 -\frac{\Lambda^2}{\tilde{m}^2+\Lambda^2}\right]} \;
\eea gives the critical $G^2$ for $B=0$, and \be b = B \;
\frac{8\pi^2}{\tilde{m}^2\ln{\left(1+\frac{\Lambda^2}{\tilde{m}^2}\right)}}
\;. \ee Details are  given in Appendix~A in Ref.\cite{042}. Note
that $B$ may be positive or negative, or more generally contains a
complex phase. The solution condition for more general cases is to
be further investigated. So, a nontrivial Majorana masses solution
is possible, or as likely as that of the case for the Dirac mass
generation.

\section{Dirac versus Majorana Masses}
While the above analysis established the holomorphic
four-superfield interaction as being capable of dynamically
generating superfield Majorana masses, the result can put the
Dirac mass generation scenario under question. After all, true
Dirac mass means a nonzero ${\mathcal M} \Phi_+ \Phi_-$  mass term
without ${\mathcal M}_+$ and ${\mathcal M}_-$. It  is important to
note that the  mass terms that arise have direct implications on
the resultant  symmetry breaking pattern. In the explicit example
of the $\Phi_+^{ai}$ and $\Phi_-^{\bar{a}j}$ multiplet illustrated
above, we can see that the ${\mathcal M}_+$ and ${\mathcal M}_-$
mass terms  and the Dirac mass term  ${\mathcal M}$ break
different parts of the symmetries in the Lagrangian. For example,
in the application to the electroweak symmetry breaking
\cite{034,042}, any Majorana mass term would be breaking the color
symmetry. A fully general analysis considering a generic mass
matrix for the $\Phi_+$ and $\Phi_-$ superfields may have to be
performed to answer the important question \cite{next} of under
what condition will the interaction give rise to any particular
symmetry breaking pattern.  However, it is interesting to see that
with the gap equation results we have so far, one can get a strong
indication of how a pure Dirac mass generation may be obtained.
The bottom line is that a split in soft masses favors Dirac mass
generation over Majorana mass generation. Let us illustrate the
story.

We note in passing that the gap equation result involves a $1/N$
approximation, with $N$ being $N_c$ for the Dirac case and  $N_r$
for the Majorana case, and there is a technical complication at
the limit where the approximation is no good or invalid. It will
be interesting to see if a gap equation can be obtained for the
case $N_r=1$. It is reasonable to think that the mass generation
and symmetry breaking mechanism still works in the case for a
sufficiently strong interaction. We will get back to discussing
the $N_r$ and $N_c$ dependence issue and its implication further
at the end  of Sec. IV.

In the discussion below, we will compare the gap equation results
for the Majorana mass generations discussed above, i.e., assuming
no off-diagonal ${\mathcal M} \Phi_+ \Phi_-$ mass term, and that
of the (pure) Dirac mass generation analysis of Ref.\cite{042}
where there is also the hidden assumption of no diagonal
(Majorana) masses ${\mathcal M}_+$ and ${\mathcal M}_-$. This is
not the full rigorous way to address the question of what would be
the resultant mass matrix of $\Phi_+$ and $\Phi_-$, but we can at
least give some insight into some qualitative aspect of the
question.

We are focusing here on the effect of the soft masses parameters on the symmetry
breaking. The more trivial effect of the $N_r$ and $N_c$ values is neglected for
the moment. To be exact, one may consider us as taking $N_r=N_c=N$ and rewriting
$G N$ as simply $G$ below.

As in the above discussions on the Majorana mass case, we focus on
the simpler case with $B=0$. Eliminating $G$ ( i.e., $G N_r$) from
the gap equations, we get the relations \be 2 |m_\pm|^2
I_1(|m_\pm|^2,\tilde{m}^2_\pm,|\eta_\pm|, \Lambda^2) =
|\eta_\mp|^2 I_2(|m_\mp|^2,\tilde{m}^2_\mp,|\eta_\mp|, \Lambda^2)
\;. \label{sol-c} \ee The integrals are given by \footnote{The
expression is formally equivalent to the earlier form given in
Ref.\cite{042}, where the $\tanh^{-1}$ function is involved,
which, straightly speaking, has a domain of definition problem.}
\bea
I_1(|m|^2, \tilde{m}^2, |\eta|, \Lambda^2) &=& \frac{1}{16\pi^2}
\left[ \frac{1}{2}(|m|^2+\tilde{m}^2)
\ln{\frac{(\Lambda^2+|m|^2+\tilde{m}^2)^2-|\eta|^2}{(|m|^2+\tilde{m}^2)^2-|\eta|^2}}
\right. \nonumber \\ && \hspace*{-.8in} \left. -|m|^2
\ln{\frac{(\Lambda^2+|m|^2)}{|m|^2}}  + \frac{|\eta|}{2}
\ln{\frac{(\Lambda^2+|m|^2+\tilde{m}^2+|\eta|)(|m|^2+\tilde{m}^2-|\eta|)}
{(\Lambda^2+|m|^2+\tilde{m}^2-|\eta|)(|m|^2+\tilde{m}^2+|\eta|)}}
\right] \;,
\nonumber \\
I_2(|m|^2, \tilde{m}^2, |\eta|, \Lambda^2)
&=&
\frac{1}{32\pi^2} \left[ \ln
\frac{(\Lambda^2+|m|^2+\tilde{m}^2)^2-|\eta|^2}{(|m|^2+\tilde{m}^2)^2-|\eta|^2}
\right. \nonumber \\
&& \hspace*{-.8in} \left. +\frac{|m|^2+ \tilde{m}^2}{|\eta|} \ln
\frac{(\Lambda^2+|m|^2+\tilde{m}^2+|\eta|)(|m|^2+\tilde{m}^2-|\eta|)}
{(\Lambda^2+|m|^2+\tilde{m}^2-|\eta|(|m|^2+\tilde{m}^2+|\eta|))}
\right]\; . \eea On the $|m|-|\eta|$ plane, the $I_2$ expression
is positive definite while the $I_1$ expression has maximum value
at the origin given by $\frac{\tilde{m}^2}{16\pi^2}
\ln{\frac{(\Lambda^2+\tilde{m}^2)}{\tilde{m}^2}}$, which implies
that $I_1$ will always be negative for $\tilde{m}^2=0$. What is
interesting then is that in the case where one of the soft masses
vanish, say, $\tilde{m}^2_+=0$, $|m_+|$ and $|\eta_-|$ will then
be forced to vanish for the above relation to be satisfied. But
the tachyonic bound for $|\eta_+|$ [$|\eta_\pm|<
(|m|^2+\tilde{m}^2_\pm)/2$ for the Majorana case] will then force
it to vanish, hence, giving also $|m_-|=0$. That is independent of
the value of the coupling.

The argument above shows there will be no nontrivial Majorana
masses generated in the case in which one of the soft masses
vanish. Vanishing, common, soft masses give no dynamical Dirac
mass either \cite{042}. In general, it is easy to see that smaller
soft supersymmetry breaking mass disfavors the dynamical mass
generations, requiring a strong coupling $G$ to achieve it, as
also explicitly illustrated by the expression for the $G$
threshold given in Ref.\cite{042}. However, when there is a
splitting between the soft masses of the two superfields, there is
a crucial difference between the Majorana mass and Dirac mass
cases. While the supersymmetric part $m$ of each Majorana mass is
directly sensitive to the vanishing of the corresponding soft
mass, the Dirac mass result is more sensitive to the average of
the two soft masses \cite{042}. We have seen that when one $m$
vanishes, it implies that all Majorana mass parameters vanish, at
least for $B=0$. We will show explicitly in the next section that
having one vanishing soft mass does not adversely affect the
dynamical generation of Dirac mass. Having one small soft mass
decreases the possibility of Majorana mass generation, pushing up
the threshold coupling, but has limited effect on the Dirac mass
generation.

\section{Dirac mass generation with one of the soft masses vanishing}
The gap equation for the case of Dirac mass generation with two
different soft masses has essentially been given in Ref.\cite{042}
(see Appendix B). We have (with  $G$ in place of $G N_c$) \bea m
&=& \frac{\bar{\eta} G}{2}\;
I'_2(|m|^2,\tilde{m}^2_+,\tilde{m}^2_-,|\eta|, \Lambda^2)\;,
\nonumber \\
\eta &=&  \bar{m} G \;
I'_1(|m|^2,\tilde{m}^2_+,\tilde{m}^2_-,|\eta|, \Lambda^2) -
\frac{\bar{\eta} G B}{2} \;
I'_2(|m|^2,\tilde{m}^2_+,\tilde{m}^2_-,|\eta|, \Lambda^2) \;, \eea
where $I'_1$ and $I'_2$ are the loop integrals. The expressions of
the integrals as given in Ref.\cite{042}, however, have some typos
and have not been written in the best form. The integral should be
exactly \small \bea I'_1 (|m|^2, \tilde{m}_+^2,\tilde{m}_-^2,
|\eta|,\Lambda^2) &=& 2\!\!  \int\!\! \frac{d^4k}{(2\pi)^4}
\frac{(\frac{\tilde{m}_+^2+\tilde{m}_-^2}{2} ) \left(
k^2+|m|^2+\frac{\tilde{m}_+^2+\tilde{m}_-^2}{2} \right) - \left(
\frac{\tilde{m}_+^2-\tilde{m}_-^2}{2} \right)^2-|\eta|^2}
{(k^2+|m|^2) \left[ \left(
k^2+|m|^2+\frac{\tilde{m}_+^2+\tilde{m}_-^2}{2} \right)^2 -
\left(\frac{\tilde{m}_+^2-\tilde{m}_-^2}{2} \right)^2 -|\eta|^2
\right]} \;,
\nonumber \\
I'_2(|m|^2, \tilde{m}_+^2,\tilde{m}_-^2, |\eta|, \Lambda^2)
&=&
\int\frac{d^4k}{(2\pi)^4}\frac{1}
{ \left[ \left( k^2+|m|^2+\frac{\tilde{m}_+^2+\tilde{m}_-^2}{2} \right)^2
- \left(\frac{\tilde{m}_+^2-\tilde{m}_-^2}{2} \right)^2-|\eta|^2 \right]} \;.
\eea \normalsize
What we failed to highlight in that paper are  the relations
\bea
I'_1 (|m|^2, \tilde{m}_+^2,\tilde{m}_-^2, |\eta|,\Lambda^2)
&=&
I_1 (|m|^2, \tilde{m}^2_{av}, |\eta'|,\Lambda^2)  \;,
\nonumber \\
I'_2 (|m|^2, \tilde{m}_+^2,\tilde{m}_-^2, |\eta|, \Lambda^2) &=&
I_2 (|m|^2, \tilde{m}^2_{av}, |\eta'|,\Lambda^2)   \;, \eea where
\bea \tilde{m}^2_{av} &=& \frac{\tilde{m}_+^2+\tilde{m}_-^2}{2}
\eea is the average value of the soft masses and \bea |\eta'|=
\sqrt{|\eta|^2 + \left(\frac{\tilde{m}_+^2-\tilde{m}_-^2}{2}
\right)^2} \;. \eea The relations are actually easy to appreciate
from the physics point of view by comparing how the various
parameters go into the (scalar) mass eigenvalues in the cases with
different or identical soft masses.

From the above, one can easily use the properties of the $I_1$ and
$I_2$ integrals to look at the case of Dirac mass generation with
only one of the soft masses vanishing. For example, one can check
that since $I_1$ increases as $|m|$ and $|\eta'|$ decrease, it
attains a maximum at $|m|=0$ and $|\eta'|$ minimum of $\left|
\frac{\tilde{m}_+^2-\tilde{m}_-^2}{2} \right|$ which corresponds
to $|\eta|=0$, and the maximum value is given by exactly the same
expression as before, namely, $\frac{2\tilde{m}^2_{av}}{16\pi^2}
\ln{\frac{\Lambda^2+2\tilde{m}^2_{av}}{2\tilde{m}^2_{av}}}$. So,
twice the average soft mass here, which equals the single nonzero
soft mass, plays the role of the single soft mass relevant for the
particular individual Majorana mass equation or the universal soft
mass case for the Majorana as well as Dirac mass generation
analysis. The simple conclusion is that having one vanishing soft
mass does not kill the Dirac mass generation as it does to the
Majorana mass generations.

\begin{figure}[!t]
\begin{center}
{\epsfig{figure=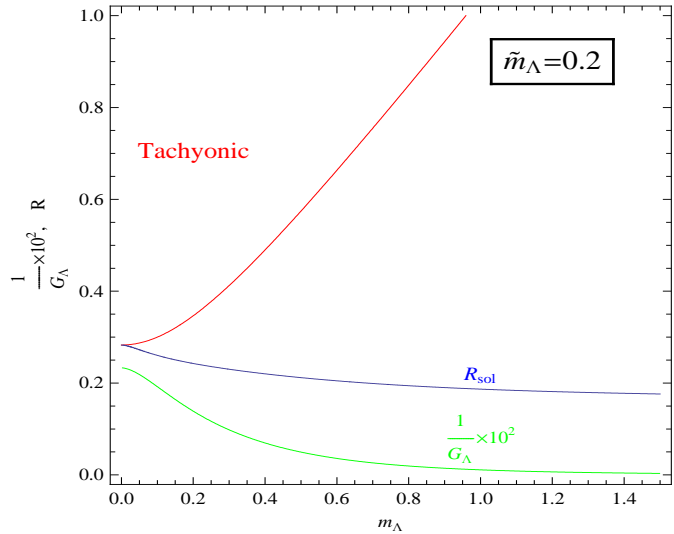,height=8.0cm,width=10.0cm}}
\end{center}
\caption{\small Illustrative numerical plot of the solution curve
on the $R_{\!\ssc \Lambda}$-$m_{\!\ssc \Lambda}$ plane. $R_{\!\ssc
\Lambda}$ and $m_{\!\ssc \Lambda}$ are corresponding dimensionless
parameters for $R=|\eta|/|m|$  and $|m|$ normalized to the basic
(cutoff) scale $\Lambda$.   Likewise, $\tilde{m}^2_{\!\ssc
\Lambda}=\tilde{m}^2_{av}/{\Lambda}^2$. The value of $1/G_{\!\ssc
\Lambda}$ versus $m_{\!\ssc \Lambda}$, $G_{\!\ssc \Lambda}=|G|
{\Lambda}$ being the normalized coupling, is also given. }
\vspace*{.2in} \hrule
\end{figure}

A further look at the solution curve as given by Eq.(\ref{sol-c})
of this Dirac case, $\frac{\bar{\eta}}{2m} I_2=
\frac{\bar{m}}{\eta} I_1$ ($=\frac{1}{G}$), on the plane of
$R=|\eta|/|m|$   versus $|m|$ (for fixed $\Lambda$ and
$\tilde{m}^2_{av}$) is particularly illustrative.  A case example
is plotted in Fig.~1.  It should be noted that $R$ has mass
dimension one while $\eta$ has mass dimension two. In general,
 $R$ decreases monotonically with $|m|$, slowing down to an
asymptotical constant value at the $|m| \to \infty$ limit. On the
physics side, it is sensible to restrict all mass parameters not
to go beyond the scale of order $\Lambda$, which is the model
cutoff scale. The asymptotical analysis helps though to show the
mathematical features. A careful calculation of the limiting
expressions for $I_1$ and $I_2$ gives the asymptotic value as
$R_{\infty}=\sqrt{\frac{2\tilde{m}^2_{av}}{3}}$, a constant. For
the $|m| \to 0$ limit, the limiting expressions for $I_1$ and
$I_2$ give $I_1=\tilde{m}^2_{av} I_2$, hence, the constant
$R_o=\sqrt{2\tilde{m}^2_{av}}$. The latter is right at the
boundary of the inadmissible tachyonic region (for lighter scalar
mass eigenstate from $A_\pm$). The bound itself is given by
$R_t=\sqrt{2\tilde{m}^2_{av}+|m|^2}$ increasing with $|m|$. The
solution curve is safely below the tachyonic bound. One can also
get the asymptotic expressions for the coupling $G$, or rather
$|G|$. We have
%$|G_{\infty}|= \frac{32\pi^2}{\sqrt{3}} \frac{|m|^4}{ \tilde{m}_{av}\Lambda^4}$,
$|G_{\infty}|= {32{\sqrt{6}}\pi^2}{|m|^4} /{ \tilde{m}_{av}\Lambda^4}$,
and the threshold coupling $|G_{o}| =16 \sqrt{2} \pi^2
\left[\tilde{m}_{av} \ln(1+\frac{\Lambda}{2\tilde{m}^2_{av}}) \right]^{-1}$.

Before we conclude, let us look a bit more into the question about
the $N_c$ and $N_r$ values. For that matter, let us take the case
of identical soft supersymmetry breaking masses. What is really
relevant here is the number of (color) states involved in the
$\Phi_+^{ai}$ and $\Phi_-^{\bar{a}j}$ pair we simply denote by
$N_c$ and the numbers of  states involved in
$\Phi_+^{ai}\Phi_+^{bi}$ and  $\Phi_-^{aj}\Phi_-^{bj}$. $G N_c$ is
the is the "coupling" parameter that shows up in the gap equation
for Dirac mass ${\mathcal M}$ while  the last two, say, $N_+$ and
$N_-$, show up  in the place of $N_c$ in the gap equations for
${\mathcal M}_-$ and ${\mathcal M}_+$, respectively. Note that
$N_+$ and $N_-$ may in general be different, say involving
different symmetries like $SO(N_+)$ and $SO(N_-)$, or in different
representations of one $SO(N_r)$. For $N_+ \ne N_-$, the Majorana
mass generation will be constrained by the smaller $N_{\pm}$
value, as one cannot have a nontrivial solution to only one of the
two masses. But the mass values will have the $N_+$ and $N_-$
ratio. In both the Dirac and Majorana cases, it is clear that as
the strong coupling condition is on the $G N$ product, a large $N$
makes the dynamical generation of the relevant mass term easier. A
coupling value $G$ large enough, for instance, for $G N_c$ to be
above the critical limit but have $G N_r$ below the limit will be
expected to give pure Dirac mass. The situation reverses with $N_r
> N_c$.

The case for $N=1$ is special in the sense that the $1/N$
approximation behind the gap equation derivation is completely not
justified. More effort may be needed to obtain  a useful gap
equation for the case. Even so, an analysis equivalent to taking
the exact gap equation here is common in the literature, starting
from the quenched planar approximation of QED by Bardeen {\it
et.al.} \cite{Bll}. Qualitatively, it is reasonable to believe
that the dynamical mechanism still works for large enough $G$. And
at least in the case that the dynamical mechanism works, the
interesting feature of Dirac mass generation being sensitive
mostly to the average of the two soft masses while Majorana masses
to the smaller soft mass between the two will very likely survive.

\section{Conclusions}
We show that the HSNJL model is capable of dynamically generating
Majorana masses for the two superfields involved. This is an
important alternative to the Dirac mass generation analyzed
earlier. The general question of under what condition a particular
mass pattern, or mass matrices, for the two superfields will
result becomes an important question to address theoretically. It
also has a phenomenological implication on whether there is a
successful application of the model to electroweak symmetry
breaking. We give here a first answer to the complicated question,
that a splitting in the  input soft supersymmetry breaking masses
favors Dirac over Majorana mass. In particular, in the limit where
one soft mass vanishes, nontrivial Majorana mass is not possible
while Dirac mass can still be generated. Hence, one expects that a
strong coupling within a particular range dictated by the
different soft masses' values will give a symmetry breaking answer
that corresponds to the pure Dirac mass.

In the case of the application to the electroweak symmetry
breaking \cite{034,042}, the model has a four-superfield
interaction involving three different gauge multiplets and hence
three soft masses of the top and bottom squark sector. It has
already been noted that a split in the soft masses between the two
sectors at least will be needed to give the phenomenologically
required top and bottom mass ratio. It is encouraging to see that
the kind of mass splitting also disfavors the generation of
Majorana masses for the quark superfield multiplets, which of
course will break color and electric charge symmetry. We note also
that the strong QCD attraction will also play an important role
there favoring color singlet vacuum condensates.
Full details of all those await further analysis. The most tricky part
is the fact that $N_r=1$ is involved. However, we
consider it reasonable to claim, from what we have been able to
establish so far, that the HSNJL stands qualitatively viable as a
model for electroweak symmetry breaking. We hope to be able to
present in a future publication full quantitative results.

Finally, we want to comment on the perspective of using an
effective field theory approach to look at the symmetry breaking.
Assuming a certain auxiliary Higgs (super)field formation as a
composite, one can write down the effective field theory on the
kind of model \cite{BE,034,042}. An (infinite) wave functional
renormalization makes the Higgs (super)field dynamical at low
energy. One can apply an analysis of the effective field theory to
find the symmetry breaking solution. This was what was done in
Ref.\cite{BE} for the case of the dimension six four-superfield
interaction. The gap equation approach accomplished in
Ref.\cite{042}, in comparison, can be considered as deriving
instead of assuming the composite formation. It gives the
nontrivial mass, hence, symmetry breaking solution, without
putting in any composite structure. The composite formation as
Higgs includes its role in having mass-generating nontrivial
vacuum. For the old model, we  achieve that and verify the
explicit results of Ref.\cite{BE}. For the HSNJL model, we derive
the result of composite Higgs formation from a Dirac pair assuming
no composite of Majorana pairing, and vice versa. Assuming a
particular form of composite formation, an effective field theory
analysis is in principle capable of determining the existence of
nontrivial vacuum and the corresponding condition on model
parameters. Such results from Ref.\cite{BE} agree with those we
obtain from a gap equation analysis \cite{042}. However, such
analysis for the HSNJL model has not been available. More
importantly, the approach is not considered more powerful in
resolving the question of under what condition the model  will
produce the Dirac or Majorana pair composite as the symmetry
breaking Higgs. Actually, the composite can be something in
between the Dirac or pure Majorana cases considered. The
nontrivial fermion/superfield masses generated may have, in
principle, mass eigenstates as any linear combination of $\Phi_+$
and $\Phi_-$. At this point, we actually do not see a way to
resolve the problem within the effective field theory approach.
Our gap equation approach, in principle, certainly can answer the
question. We  obtain the full gap equation for the generic
superfield mass matrix \cite{next}. It is taking more hard work to
extract explicit solution information though.

\acknowledgments Y.-M.D.,G.F., and O.K. are partially supported by
research Grant No. NSC 99- 2112-M-008-003-MY3, and G.F. is further
supported by Grants No. NSC 100-2811-M-008-036 and No. NSC
101-2811-M-008-022 of the National Science Council of Taiwan.
D.-W.J. is supported by NRF Research Grant No. 2012R1A2A1A01006053
of Korea.
%\appendix

\end{document}